\documentclass[a4paper,showpacs,preprintnumbers,amsmath,amssymb,pre,twocolumn,superscriptaddress]{revtex4}
\usepackage[T1]{fontenc}
\usepackage{pgf}
\usepackage{xcolor}
\usepackage[lined,boxed,commentsnumbered]{algorithm2e}
\usepackage{verbatim}
\usepackage{amsmath}
\usepackage{amssymb}
\usepackage{amsfonts}
\usepackage{subfig}
\usepackage{array}
\usepackage{hyperref}
\usepackage{bm,bbm}

\newcommand{\zero}{0}
\newcommand{\one}{1}
\newcommand{\eg}{\emph{e.g.\/}}

\newcommand{\tr}{\mathrm{tr}}
\newcommand{\Tr}{\tr}

\newcommand{\ket}[1]{\ensuremath{|#1\rangle}}
\newcommand{\bra}[1]{\ensuremath{\langle#1|}}
\newcommand{\ketbra}[2]{\ensuremath{\ket{#1}\bra{#2}}}

\newcommand{\1}{\mathbbm{1}}

\newcommand{\h}{\mathcal{H}}

\newcommand{\J}{\ensuremath{\mathcal{J}}}
\begin{document}

\title{Enhancing pseudo-telepathy in the Magic Square game}

\author{{\L}ukasz Pawela}
\email{lpawela@iitis.pl}
\author{Piotr Gawron}
\email{gawron@iitis.pl}
\affiliation{Institute of Theoretical and Applied Informatics, Polish Academy
of Sciences, Ba{\l}tycka 5, 44-100 Gliwice, Poland}
\author{Zbigniew Pucha{\l}a}
\email{z.puchala@iitis.pl}
\affiliation{Institute of Theoretical and Applied Informatics, Polish Academy
of Sciences, Ba{\l}tycka 5, 44-100 Gliwice, Poland}
\affiliation{Institute of Physics, Jagiellonian University, Reymonta 4, 30-059 Krak{\'o}w, Poland}
\author{Jan S{\l}adkowski}
\email{jan.sladkowski@us.edu.pl}
\affiliation{Institute of Physics, University of Silesia,
Uniwersytecka 4, 40-007 Katowice, Poland}

\date{19/IV/2013}


\begin{abstract}
We study the possibility of reversing an action of a quantum channel. Our principal objective
is to find a specific channel that reverses as accurately as possible an action
of a given quantum channel. To achieve this goal we use semidefinite
programming. We show the benefits of our method using the quantum pseudo-telepathy
Magic Square game with noise. Our strategy is to move the pseudo-telepathy
region to higher noise values. We show that it is possible to reverse the action
of a noise channel using semidefinite programming.
\end{abstract}

\maketitle

\section{Introduction}
Quantum game theory is an interdisciplinary field that combines game theory
and quantum information. It lies at the crossroads of physics, quantum
information processing, computer and natural sciences. Various quantizations of
games were presented by different authors \cite{eisert1999quantum,
piotrowski2003invitation, flitney2002quantum, frackiewicz2012quantum,
schmidt2012quantum}.

Quantum pseudo-telepathy games~\cite{brassard2005quantum} form a subclass of
quantum games. A game belongs to the pseudo-telepathy class providing that there are no
winning strategies for classical players, but a winning strategy can be found
if the players share a sufficient amount of entanglement. In these games
quantum players can accomplish tasks that are unfeasible for their classical
counterparts. It has been shown~\cite{noisy_square} that noise in a quantum
channel can decrease the probability of winning the Magic Square game even
below the classical threshold.

Noise is an unavoidable ingredient of a quantum system. Therefore its thorough
investigation is a fundamental issue in quantum information processing. Quantum
game theory has several potential applications (e.g quantum auctions
\cite{piotrowski2008quantum}) that may be hindered by noise effects. Our
previous investigation of quantum noise effects in quantum games
\cite{noisy_square,gawron2010noisy,miszczak2011qubit}, and quantum algorithms
performance \cite{gawron2012noise} revealed several interesting issues that act
as an incentive of the present work. The tools developed in this work can be
used to analyse the behaviour of quantum channels in other settings.

This paper is organized as follows. In Section~\ref{sec:motivation} we motivate
our research. In Section~\ref{sec:square} we present the  Magic Square game. In
Section~\ref{sec:quantu-channels} we recall essential facts about quantum
channels. In Section~\ref{sec:noise} we discuss noisy quantum channels. In
Section~\ref{sec:model} we introduce a~method of reversing an action of
a~channel. Section~\ref{sec:results} contains our results and their discussion.
Finally, in Section~\ref{sec:conc} conclusions are drawn.

\section{Motivation}\label{sec:motivation}
The motivation to study the Magic Square game and pseudo-telepathy games in
general is that their physical implementation could provide convincing, even to
a layperson, demonstration that the physical world is not local realistic. By
\emph{local} we mean that no action performed at some location X can have an
effect on some remote location Y in a time shorter then that required by light
to travel from X to Y. \emph{Realistic} means that a measurement can only
reveal elements of reality that are already present in the
system~\cite{brassard2005quantum}.

Given a pseudo-telepathy game, one can implement a quantum winning strategy for
this game \cite{brassard2005quantum}. In an ideal case, the experiment should
involve a significant number of rounds of the game. The experiment should be
continued until either the players lose a single round or the players win
such a great number of rounds, that it would be nearly impossible if they were
using a classical strategy.

In the particular case of the magic square game the classical strategy allows
the players to achieve the success rate no larger than $\frac{8}{9}$. In theory, the
success rate of the quantum strategy is equal to one. But any physical
implementations of a quantum protocol cannot be perfect because it is subject to
noise.

In particular, the players, Alice and Bob, must be  so far away from each other
that the time between the question and their  respective answers is shorter
than the time required by light to travel between their locations. This set-up
involves sending parts of an entangled quantum state to two remote locations.
Sending qubits through a channel will inevitably add noise to the system. Our
aim is to counteract this noise.
In this paper we focus on the destructive aspects of the process of transmission
of a qubit through a noisy separable quantum channel and introduce a scheme
that allows  the partial reversion of the channel action. This reversal gives
rise to the players'  success rate above the classical limit of $8/9$ for some
parameters of noisy channels. Our scheme for reversing an action of a noisy
channel may prove valuable in future experimental set-ups of such games.

\section{Magic Square game}\label{sec:square}
The magic square is a $3 \times 3$ matrix filled with numbers 0 or 1 so that the
sum of entries in each row is even and the sum of entries in each column is odd.
Although such a matrix cannot exist (see Fig.~\ref{fig:impossible}) one can
consider the following game.

\begin{figure}[!htp]
	\begin{tabular}{|>{\centering\arraybackslash}m{1em}| >{\centering\arraybackslash}m{1em}|>{\centering\arraybackslash}m{1em}|}
	\hline
	\one & \one & \zero \\ \hline
	\one & \zero & \one \\ \hline
	\one & \zero & ? \\ \hline
	\end{tabular}
	\caption{An illustrative filling of the magic square with numbers 0 and 1. The 
	question mark shows that it is not possible to put a number in the last 
	field and satisfy both conditions of the game.}\label{fig:impossible}
\end{figure}

The game setup is as follows. There are two players: Alice and Bob. Alice is
given a row, Bob is given a column. Alice has to give
the entries for a row and Bob has to give entries for a column so that the
parity conditions are met.
Winning condition is that the players' entries at the intersection must agree.
Alice and Bob can prepare a strategy but they are not allowed
to communicate during the game.

There exists a (classical) strategy that guarantees the winning probability of
$\frac{8}{9}$. If the parties are allowed to share a quantum state they can
achieve probability of success equal to one \cite{brassard2005quantum}.

In the quantum version of this game~\cite{mermin, aravind2004quantum} Alice and
Bob are allowed to share an entangled quantum state. The winning strategy is
following. Alice and Bob share an entangled state:
\begin{equation}
\ket{\psi} = \frac{1}{2}\left( \ket{0011} + \ket{1100} - \ket{0110} - \ket{1001} \right)
\end{equation}
and apply local unitary operators forming operator $A_i\otimes B_j$, where

\begin{tabular}{l@{=}ll@{=}l}
$A_1$ &
$
\frac{1}{\sqrt{2}}
\left(
\begin{smallmatrix}
i & 0 & 0 & 1\\
0 &-i & 1 & 0\\
0 & i & 1 & 0\\
1 & 0 & 0 & i
\end{smallmatrix}
\right)
,
$ &
$A_2$ &
$
\frac{1}{2}
\left(
\begin{smallmatrix}
i & 1 & 1 & i\\
-i & 1 & -1 & i\\
i & 1 & -1 & -i\\
-i & 1 & 1 & -i
\end{smallmatrix}
\right)
,
$ \\
$A_3$ &
$
\frac{1}{2}
\left(
\begin{smallmatrix}
-1 & -1 & -1 & 1\\
1 & 1 & -1 & 1\\
1 & -1 & 1 & 1\\
1 & -1 & -1 & -1
\end{smallmatrix}
\right)
,
$ &
$B_1$ &
$
\frac{1}{2}
\left(
\begin{smallmatrix}
i & -i & 1 & 1\\
-i & -i & 1 & -1\\
1 & 1 & -i & i\\
-i & i & 1 & 1
\end{smallmatrix}
\right)
,
$ \\
$B_2$ &
$
\frac{1}{2}
\left(
\begin{smallmatrix}
-1 & i & 1 & i\\
1 & i & 1 & -i\\
1 & -i & 1 & i\\
-1 & -i & 1 & -i
\end{smallmatrix}
\right)
,
$ &
$B_3$&
$
\frac{1}{\sqrt{2}}
\left(
\begin{smallmatrix}
1 & 0 & 0 & 1\\
-1 & 0 & 0 & 1\\
0 & 1 & 1 & 0\\
0 & 1 & -1 & 0
\end{smallmatrix}
\right).
$
\end{tabular}

Indices $i$ and $j$ label rows and columns of the magic square.
The state of this scheme before measurement is
\begin{equation}\label{eq:game-final-state}
\rho_f=(A_i\otimes B_j)\,
\ket{\psi}\bra{\psi}\,(A_i^\dagger\otimes B_j^\dagger).
\end{equation}
The final
step of the game consists of the measurement in the computational basis.

In \cite{noisy_square}, the situation where the initial  state $\ket{\psi}$ is
corrupted by the noise was investigated. Therefore, Eq.~\ref{eq:game-final-state}
is transformed into
\begin{equation}\label{eq:game-final-state-noisy}
\rho_f=(A_i\otimes B_j)\,
\Phi_\alpha(\ket{\psi}\bra{\psi})
\,(A_i^\dagger\otimes B_j^\dagger),
\end{equation}
where $\Phi_\alpha$ denotes one-parameter family of noisy quantum channels.

In such a case it is justified to inquire what is the mean probability of Alice and
Bob's success given the amount of noise introduced by channel
$\Phi_\alpha$. The mean probability $p(\alpha)$ of measuring the outcome yielding
success in the
state $\rho_f $ is given by
\begin{equation}
p(\alpha)=\frac{1}{9}
\sum\limits_{i,j=1}^{3}
\sum_{\xi\in\mathcal{S}_{ij}}
\Tr{\rho_f}
{\ketbra{\xi}{\xi}},
\end{equation}
where $\mathcal{S}_{ij}$ is the set of right answers for the column and row $ij$ (Tab.~\ref{tbl:msg:states}).

{
\renewcommand{\baselinestretch}{1}
\renewcommand{\arraystretch}{1}
\renewcommand{\tabcolsep}{0.4mm}
\begin{table}[h]
\begin{center}
\begin{footnotesize}
\begin{tabular}{c|cccccccccccccccc}
 & 
 ${0}$&${1}$&${2}$&${3}$&${4}$&${5}$&${6}$&${7}$
 &${8}$&${9}$&${10}$&${11}$&${12}$&${13}$&${14}$&
 ${15}$\\
\hline
$\mathcal{S}_{11}$&+&+&-&-&+&+&-&-&-&-&+&+&-&-&+&+\\
$\mathcal{S}_{12}$&+&+&-&-&-&-&+&+&+&+&-&-&-&-&+&+\\
$\mathcal{S}_{13}$&+&+&-&-&-&-&+&+&-&-&+&+&+&+&-&-\\
$\mathcal{S}_{21}$&+&-&+&-&+&-&+&-&-&+&-&+&-&+&-&+\\
$\mathcal{S}_{22}$&+&-&+&-&-&+&-&+&+&-&+&-&-&+&-&+\\
$\mathcal{S}_{23}$&+&-&+&-&-&+&-&+&-&+&-&+&+&-&+&-\\
$\mathcal{S}_{31}$&-&+&+&-&-&+&+&-&+&-&-&+&+&-&-&+\\
$\mathcal{S}_{32}$&-&+&+&-&+&-&-&+&-&+&+&-&+&-&-&+\\
$\mathcal{S}_{33}$&-&+&+&-&+&-&-&+&+&-&-&+&-&+&+&-
\end{tabular}
\end{footnotesize}
\end{center}
\caption{Sets $\mathcal{S}_{ij}$ --- plus sign (+) indicates that the given element belongs to the set, minus (-) sign indicates that the element does not belong to the set.}
\label{tbl:msg:states}
\end{table}
}

A winning strategy exists for noiseless channels. In the case of noisy channel,
the same strategy gives a higher probability of winning than in the classical
case for low noise channels \cite{noisy_square}. The objective of this work is
to find local channels that partially reverse the action of the noise and
therefore extends the pseudo-telepathy to channels with higher noise. In order
to achieve this, Eq.~\ref{eq:game-final-state-noisy} is transformed into
\begin{equation}\label{eq:game-final-state-recovered}
\rho_f=(A_i\otimes B_j)\,
\Psi_\alpha(
\Phi_\alpha(\ket{\psi}\bra{\psi})
)
\,(A_i^\dagger\otimes B_j^\dagger),
\end{equation}
where $\Psi_\alpha$ denotes local channel with respect to Alice and Bob's
subsystems that allows to raise their probability of winning $p(\alpha)$. In
order to achieve that a~series of semi-definite optimization programs has to be
numerically solved.

\section{Quantum channels}\label{sec:quantu-channels}
In the most general case, the evolution of a quantum system can be described
using the notion of a \emph{quantum
channel}~\cite{BZ2006,NC2000,puchala2011experimentally}. A quantum channel is a
mapping acting on density operators $\rho \in D(\h)$, i.e., operators where
$\rho \geq 0$ and $\Tr(\rho)=1$ on a Hilbert space $\h_1$ and transforming
them into operators on a another Hilbert space $\h_2$. Thus
\begin{equation}
	\Phi: L(\h_1) \rightarrow L(\h_2),
\end{equation}
where $L(\h_i)$ denotes the set of linear operators on $\h_i$. To form a
proper quantum channel, the mapping $\Phi$ must satisfy the following
restrictions:
\begin{enumerate}
	\item $\Phi$ must be \emph{trace-preserving}, that is
	$\tr(\Phi(\rho))=\tr(\rho)$,
	\item $\Phi$ must be \emph{completely positive}, that is $\Phi\otimes
	\1_{L(\h_3)}$ is a positive mapping, thus
	\begin{equation}
		(\Phi\otimes \1_{L(\h_3)})(\rho)\in D(\h_2\otimes\h_3),
	\end{equation}
	for every choice of $\rho\in D(\h_1\otimes\h_3)$ and every choice of
	finite-dimensional Hilbert space $\h_3$, where $\1_{L(\h_3)}$ is
	an identity channel on the space $L(\h_3)$.
\end{enumerate}

The notion of a \emph{product} quantum channel is introduced as follows~\cite{watrous}. For any choice of quantum channels that satisfy
\begin{equation}
\Phi_1: L(\h_1^1) \rightarrow L(\h_2^1),\ldots, \Phi_N: L(\h_1^N) \rightarrow L(\h_2^N),
\end{equation}
we define a linear mapping
\begin{equation}
	\Phi_1 \otimes \ldots \otimes \Phi_N: L(\h_1^1\otimes \ldots \otimes\h_1^N) \rightarrow L(\h_2^1\otimes \ldots \otimes\h_2^N),
\end{equation}
to be the unique mapping that satisfies the equation
\begin{equation}
(\Phi_1 \otimes \ldots \otimes \Phi_N)(A_1\otimes\ldots\otimes A_N) = \Phi_1(A_1) \otimes\ldots \otimes \Phi_N(A_N),
\end{equation}
for all operators $A_1\in L(\h_1^1),\ldots, A_N \in L(\h_1^N)$.

Many different representations of quantum channels can be chosen, depending on
the application. Among these are the Jamio{\l}kowski representation, the Kraus
representation and the Stinespring representation. These three representations
will be used throughout this paper.

The Jamio{\l}kowski representation of a quantum channel $\Phi$ is given by
\begin{equation}
	\J(\Phi) = \sum_{a,b} \Phi(E_{a,b})\otimes E_{a,b},
\end{equation}
where $E_{a,b} \in L(\h_1)$ are operators with all entries equal to zero, except the entry $a,b$ equal to one. From this definition, it is straightforward to observe that $\J(\Phi) \in L(\h_2 \otimes \h_1)$. By the Choi's~\cite{BZ2006} theorem a channel is completely positive if and only if $\J(\Phi) \geq 0$. It is trace-preserving if and only if
\begin{equation}
	\Tr_{\h_2}(\J(\Phi)) = \1_{\h_1}.
\end{equation}
Finally, the action of a quantum channel in the Jmio{\l}kowski representation is given by
\begin{equation}
	\Phi(\rho) = \Tr_{\h_1}(\J(\Phi)(\1_{\h_2} \otimes \rho^T)).
\end{equation}

The Kraus representation of a quantum channel is given by a set of operators $E_k \in L(\h_1, \h_2)$. The action of quantum channel $\Phi$ is given by:
\begin{equation}
\Phi(\rho)=\sum_k E_k \rho {E_k}^\dagger.
\end{equation}
This form ensures that the quantum channel is completely positive. For it to be also trace-preserving we need to impose the following constraint on the Kraus operators
\begin{equation}
	\sum_k {E_k}^\dagger E_k=\1_{\h_1}.
\end{equation}

Finally, given a mapping $\Phi: L(\h_1) \rightarrow L(\h_2)$ let us take
another Hilbert space $\h_3$ such that
$\textnormal{dim}(\h_3) = \textnormal{rank}(J(\Phi))$ and a linear isometry
$A\in
U(\h_1,
\h_2\otimes\h_3)$. The action of a quantum channel is given by
\begin{equation}
	\Phi(\rho) = \tr_{\h_3}(A\rho A^\dagger).
\end{equation}
This representation is called the Stinespring representation of $\Phi$.

For further discussion of quantum channels see \eg~\cite{NC2000} or~\cite{watrous}.

\section{Quantum noise}\label{sec:noise}
In the literature, several one-parameter families of qubit noisy
channels are discussed~\cite{NC2000}.
For all the families of channels listed below the parameter $\alpha\in [0,1]$ represents the
amount of noise introduced by the channel. The symbols
$\sigma_x,\sigma_y,\sigma_z$ denote Pauli operators. The Kraus operators for
typical noisy channels are for
\begin{itemize}
\item depolarising channel:\\
$\left\{
\sqrt{1-\frac{3\alpha}{4}}\1
,
\sqrt{\frac{\alpha}{4}}\sigma_x
,
\sqrt{\frac{\alpha}{4}}\sigma_y,
\sqrt{\frac{\alpha}{4}}\sigma_z
\right\},$
\item amplitude damping:
$\left\{
\left(
\begin{smallmatrix}
1 & 0 \\
0 & \sqrt{1-\alpha}
\end{smallmatrix}
\right)
,
\left(
\begin{smallmatrix}
0 & \sqrt{\alpha} \\
0 & 0
\end{smallmatrix}
\right)
\right\},$
\item phase damping:
$\left\{
\left(
\begin{smallmatrix}
1 & 0 \\
0 & \sqrt{1-\alpha}
\end{smallmatrix}
\right)
,
\left(
\begin{smallmatrix}
0 & 0 \\
0 & \sqrt{\alpha}
\end{smallmatrix}
\right)
\right\},$
\item phase flip:
$\left\{
\sqrt{1-\alpha}\1,
\sqrt{\alpha}\sigma_z
\right\}
,$
\item bit flip:
$
\left\{
\sqrt{1-\alpha}\1,
\sqrt{\alpha}\sigma_x
\right\},
$
\item bit-phase flip:
$\left\{
\sqrt{1-\alpha}\1
,
\sqrt{\alpha}\sigma_y
\right\}$.
\end{itemize}

In order to apply noise operators to multiple qubits we form new set of
operators acting on a larger Hilbert space.

Given a set of $n$ one-qubit Kraus operators $\{e_k\}_{k=1}^n$ it is possible to
construct new set of $n^N$ operators $\{E_k\}_{k=1}^{n^N}$ that act on a
Hilbert space of dimension $2^N$ by taking Cartesian product of one-qubit Kraus
operators in the following way
\begin{equation}\label{equ:localchannel}
\{E_k\}_{k=1}^{n^N}=
\{e_{i_1}\otimes e_{i_2}\otimes\ldots \otimes
e_{i_N}\}_{i_1,i_2,\ldots i_N=1}^n.
\end{equation}

Application of the above to Kraus operators listed before
gives one-parameter families of local noisy channels. This form will be
used in further investigations.

\section{Reversing the action of a channel}\label{sec:model}

We propose the following scheme for reversing an action of a channel using
semidefinite programming (SDP). In our case, the most useful formulation of a
semidefinite program is as follows (after Watrous~\cite{watrous}).

A semidefinite program is a triple ($\Phi, A, B$) where $\Phi:
L(\h_1)\rightarrow L(\h_2)$ is a Hermiticity-preserving map and $A\in L(\h_1)$
and $B \in L(\h_2)$ are Hermitian operators for some choice of Hilbert spaces
$\h_1$ and $\h_2$. Two optimization problems are associated with the triple
$(\Phi, A, B)$, the primal and dual problems. We will focus our attention on the
primal problem, which has the form:
\begin{eqnarray*}
	\text{maximize:}&& \Tr(AX), \\
	\text{subject to:}&& \Phi(X)=B, \\
		&& X\ge 0.
\end{eqnarray*}

In the case of the pseudo-telepathy game, it seems appropriate to look for a
channel in a product form. This is due to the fact, that Alice and Bob are
separated and each of them must apply a local channel. To model this situation,
let us consider the Jamio{\l}kowski representations of Alice's and Bob's channels,
denoted $Y$ and $Z$ respectively. The resulting channel is given by
\begin{equation}
	T=W(Y \otimes Z)W,\label{eq:T_prod}
\end{equation}
where $W$ is an operator defined as follows
\begin{equation}
	W = \1_{\h_2^A}\otimes U \otimes\1_{\h_1^B},
\end{equation}
where $U \in L(\h_1^A \otimes \h_2^B, \h_2^B \otimes \h_1^A)$ is the swap
operation of subsystems $\h_1^A$ and $\h_2^B$, defined as
\begin{equation}
U = \sum_{i,j}\ketbra{f_j e_i}{e_i f_j},
\end{equation}
for $e_i, f_j$ being elements of orthonormal bases of $\h_1^A$ and $\h_2^B$
respectively.

Next, let us denote by $\Psi_N$ the noise channel and we put $\sigma =
\ketbra{\psi}{\psi}$. For simplicity of further calculations, let us write $\tau
= \Psi_N(\sigma)$ and $T = \J(\Psi_N)$. Consider the following maximization
criterion problem
\begin{equation}
	\textnormal{maximize:}\;\;\; \Tr(\Tr_2(W(Y \otimes Z)W(\1 \otimes \tau^T))\sigma), \label{eq:max_prod}
\end{equation}
which means we aim to find a channel that reverses the action of the noise
channel as accurately as possible. Unfortunately, a maximization criterion in
this form does not yield an SDP problem. To formulate this problem as an SDP, we
first conduct some simple calculations that allow us to rewrite the maximization
condition~\eqref{eq:max_prod} as
\begin{eqnarray}
	\textnormal{maximize:} & & \Tr((Y \otimes Z)M), \label{eq:new_max_prod}\\
	\textnormal{subject to:}& & M = W(\sigma \otimes \tau)W. \label{eq:M_prod}
\end{eqnarray}
Considering the value of $Y$ to be fixed and using the equation
$\tr((\1\otimes A)B)=\tr(A\ \tr_1(B))$, allows us to write the following SDP
\begin{eqnarray}
	\textnormal{maximize:} & & \Tr(Z\Tr_{\h_2^B, \h_1^B}(M(Y \otimes \1_{\h_2^B\otimes \h_1^B})), \label{eq:left} \nonumber\\
	\textnormal{subject to:} & & \Tr_{\h_2^B}(Z) = \1_{\h_1^B}, \\
	&& Z \geq 0. \nonumber
\end{eqnarray}
Fixing the value of $Z$ and following a similar calculation give the following SDP problem
\begin{eqnarray}
	\textnormal{maximize:} & & \Tr(Y\Tr_{\h_2^A, \h_1^A}((\1_{\h_2^A\otimes \h_1^A} \otimes Z)M), \label{eq:right} \nonumber\\
	\textnormal{subject to:} & & \Tr_{\h_2^A}(Y) = \1_{\h_1^A}, \\
	&& Y \geq 0. \nonumber
\end{eqnarray}
Now, we use the following algorithm to find an optimal channel. The algorithm in each iteration optimizes only a single part of the product channel. This algorithm was implemented using the SDPLR library \cite{sdplr_main,sdplr_help}
\begin{algorithm}
\SetKwData{Eqn}{Eqn.}\SetKwData{Z}{Z}\SetKwData{Y}{Y}\SetKwData{W}{W}
\SetKwData{mmmm}{M}
\SetKwFunction{Optimize}{SolveSDP}
\SetKwFunction{Return}{return}
\SetKwInOut{Input}{input}\SetKwInOut{Output}{output}
\Input{A random Jamio{\l}kowski matrix~\cite{bruzda2009random}, $Y$, the matrices $W$, $\sigma$, $\tau$ and the number of runs $n$}
\Output{Optimized values of the parts of the product channel}
\BlankLine
calculate \mmmm \;
\For{$i\leftarrow 1$ \KwTo $n$}{
\Z = \Optimize{\Eqn~\eqref{eq:left}, \Y, \mmmm}\;
\Y = \Optimize{\Eqn~\eqref{eq:right}, \Z, \mmmm}\;
}
\Return \W(\Y $\otimes$ \Z)\W
\caption{SDP optimization of a product channel}\label{algo:optimization}
\end{algorithm}

\section{Numerical results}\label{sec:results}
The numerical results are gathered in form of plots at the end of the paper.
Fig.~\ref{fig:res} shows the results of the optimization scheme described in
algorithm~\ref{algo:optimization}. The application of the SDP allowed us to achieve
greater winning probability for all types of noisy channels.

In the case of the flip channels the obtained results are depicted in
Figs.~\ref{fig:phase_flip},~\ref{fig:bit_flip} and~\ref{fig:bit_phase_flip}.
These plots show that it is possible to reverse the action of the noise channel
for all values of the noise parameter $\alpha$. Hence, we are able to observer
quantum pseudo-telepathy for higher noise channels. Furthermore, the use of our
optimization method results in a plot of probability of winning as a function of
the noise parameter $\alpha$ which has a shape similar to the case when we do
not try to reverse the action of a channel.

Next, we move to the depolarising channel. The results obtained in this case are
shown in Fig.~\ref{fig:depolarising}. Likewise, in this case our method also has
allowed us to achieve pseudo-telepathy for higher values of the noise parameter
$\alpha$. The details are depicted in the inset in Fig.~\ref{fig:depolarising}.
Additionally, for values of the noise parameter $\alpha \ge 0.45$ the
probability of winning the game stabilizes around 0.65, opposed to the case with
no channel action inverse, where it decreases to 0.5. Hence, we are able to retrieve
some information in the case of high noise, local depolarising channels acting
on many qubits.

Finally, we move to the damping channels. Numerical results for this case are
depicted in Figs.~\ref{fig:phase_damping} and~\ref{fig:amplitude_damping}. Also
in this case we are able to reverse the action of a noise channel and broaden
the pseudo-telepathy region. In the case of high values of the noise parameter
$\alpha$, results for the amplitude damping channel resemble those obtained for
depolarising channel, as the probability of winning stabilizes around 0.65 for
$\alpha \ge 0.45$ instead of decreasing to approximately 0.5.

\section{Conclusions}\label{sec:conc}

The principal result of this paper is a methodology of partial denoising with the
usage of local quantum channels. The presented tool can be used in the
cases in which
\begin{itemize}
\item the parameters of the noise are accessible,
\item the noisy channel is separable and acts independently on each qubit,
\item the entangled quantum state the parties use in known in advance,
\item the parties have access to quantum computers but
\item are no allowed to communicate.
\end{itemize}

We have proposed a method to reverse an action of a quantum channel using
semidefinite programming. The method allows us to find a product channel which
partially reverses a given channel. We use the following scheme to achieve this
goal. First, we fix all parts of the product, except for one, which is being
optimized. After the SDP optimization, we move on to optimize the next part of
the product channel, using the value obtained in the earlier step. We repeat
this for all parts of the product channel. We run the process a great number of times to
obtain a converging solution.

\begin{figure}[h]
	\centering
	\includegraphics{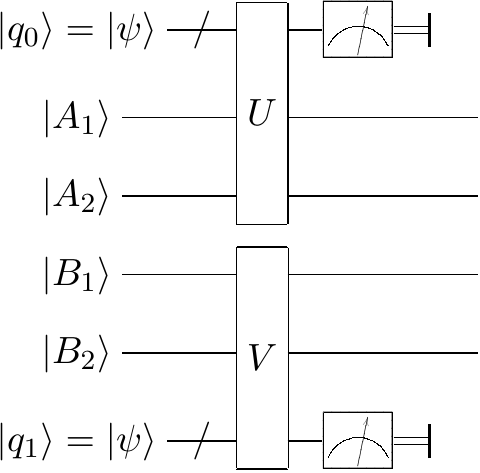}
	\caption{A quantum circuit showing the implementation of our 
	scheme. $A_i$ and $B_i$ denote Alice's and Bob's qubits. $q_0$ and $q_1$ 
	are the ancillary qubits they need to add.}\label{fig:implementation}
\end{figure}

Obtained channel may be implemented on a real physical system using the
Stinespring representation. An example of the quantum circuit implementing this
scheme is shown in Figure~\ref{fig:implementation}. Alice and Bob each add
ancillary qubits to their original ones. Then they apply a unitary operator on
their respective systems. Finally, they perform a measurement on the ancillary
qubits, leaving their starting qubits in a less noisy state.

As an example of usage of this optimization scheme we present the quantum
pseudo-telepathy magic square game. We obtained results showing an improvement
in the players' success rate in the game. Specifically, we were to broaden the
range of the noise parameter $\alpha$ for which the quantum effect occurs.

\section*{Acknowledgements}
This work was supported by the Polish National Science Centre: {\L}.~Pawela by
the grant number N~N516~481840, J.~S{\l}adkowski by the grant number
DEC-2011/01/B/ST6/07197. Work was also supported by the Polish Ministry of
Science and Higher Education: P.~Gawron under the project number IP2011~014071,
Z.~Pucha{\l}a under the project number IP2011~044271.

\bibliography{telepathy}
\bibliographystyle{apsrev4-1}

\newlength{\wdth}
\setlength{\wdth}{0.48\textwidth}
\begin{figure*}[b]
	\begin{tabular}{lr}
		\subfloat[Phase flip channel]
		{\label{fig:phase_flip}\includegraphics[width=\wdth]{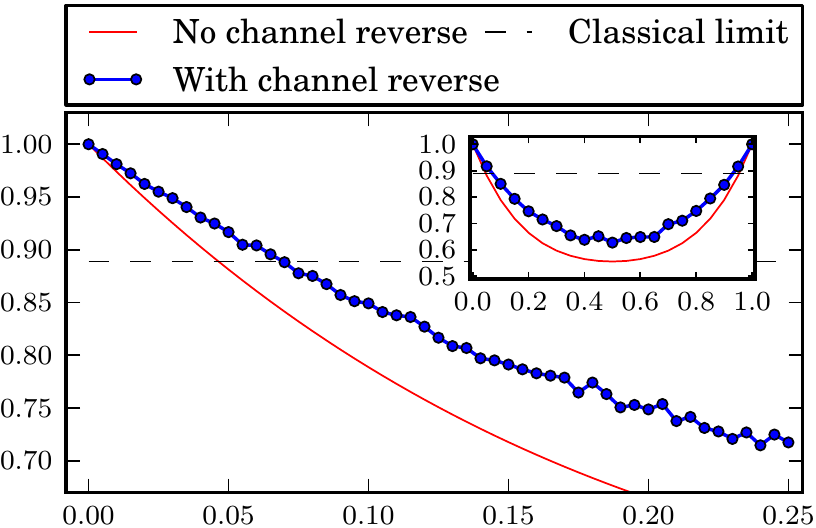}}&
		\subfloat[Bit flip channel]
		{\label{fig:bit_flip}\includegraphics[width=\wdth]{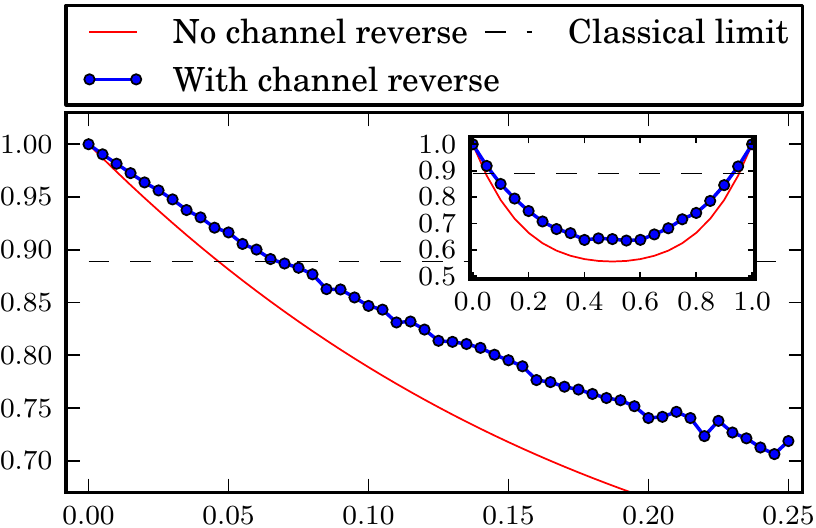}} \\
		\subfloat[Bit phase flip channel]
		{\label{fig:bit_phase_flip}\includegraphics[width=\wdth]{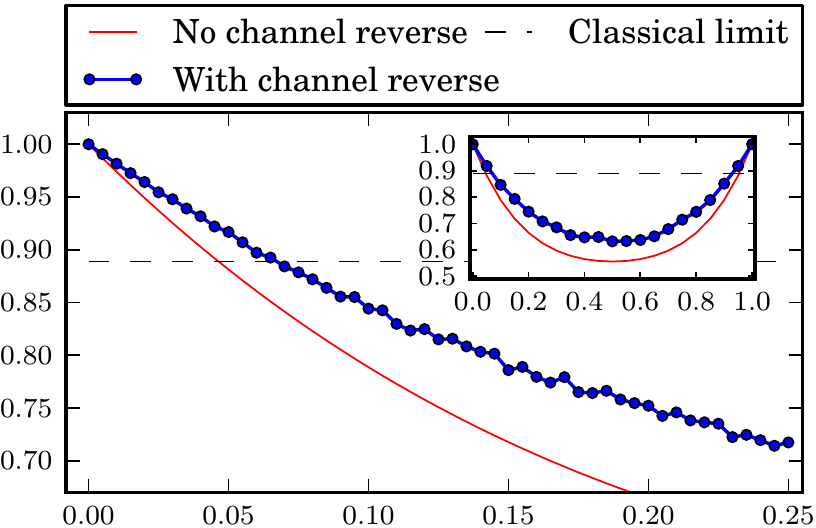}}&
		\subfloat[Depolarising channel]
		{\label{fig:depolarising}\includegraphics[width=\wdth]{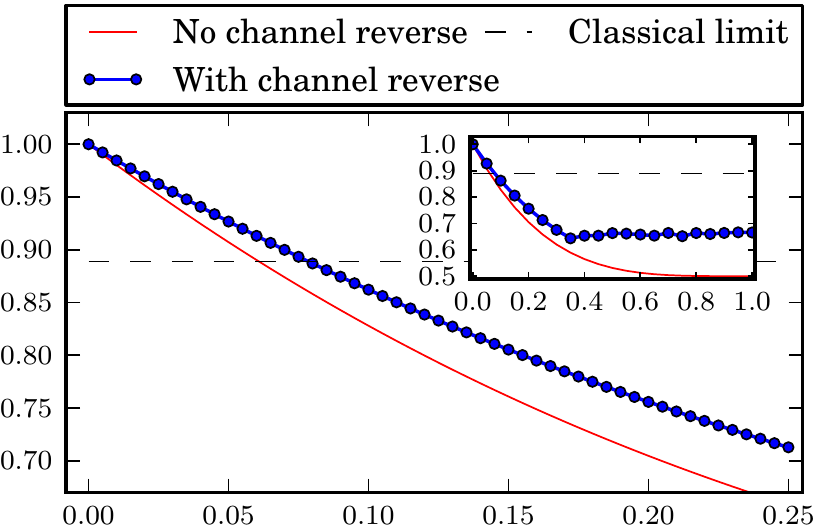}} \\
		\subfloat[Phase damping channel]
		{\label{fig:phase_damping}\includegraphics[width=\wdth]{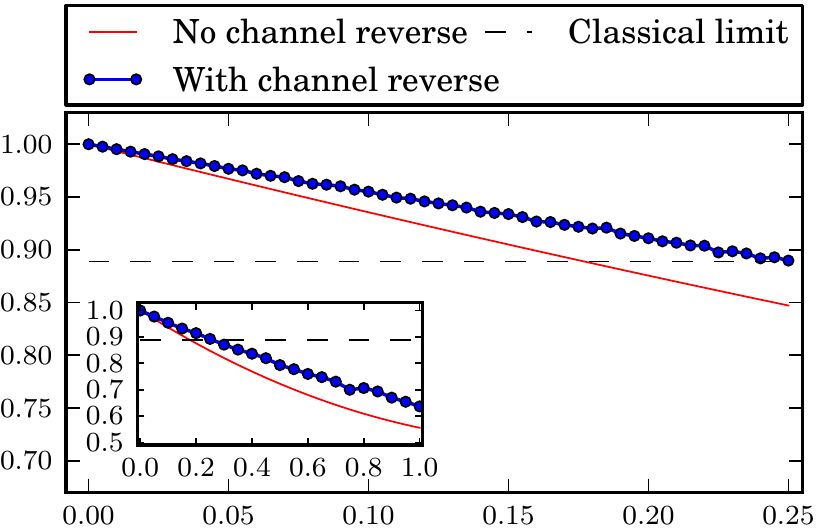}}&
		\subfloat[Amplitude damping channel]
		{\label{fig:amplitude_damping}\includegraphics[width=\wdth]{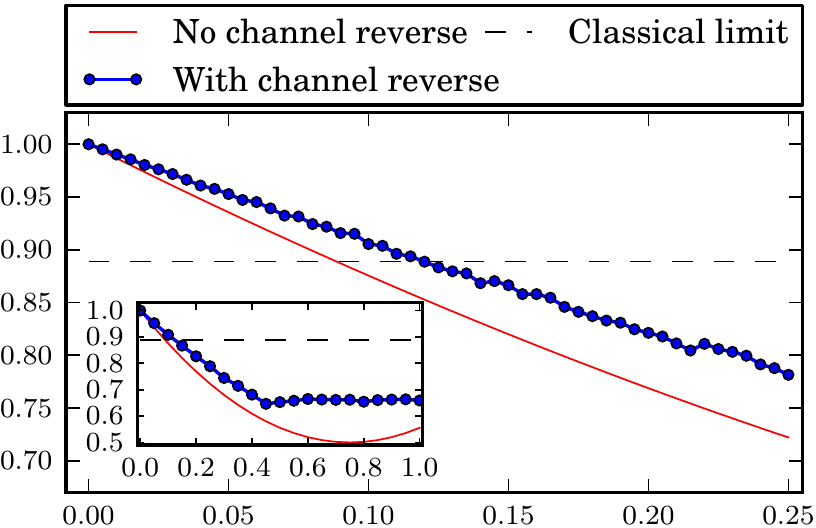}}
	\end{tabular}
	\caption{Probability of winning the pseudo-telepathy game with and without the use of our approach as a function of the noise parameter $\alpha$ for different noise channels. Lines connecting the points are eye-guides. The insets show the probability of winning for $\alpha\in[0,1]$.}\label{fig:res}
\end{figure*}

\end{document}